\documentclass[12pt]{article}
\newcommand{\nn}{\nonumber\\}
\newcommand{\calL}{\mathcal{L}}
\newcommand{\calD}{\mathcal{D}}
\newcommand{\diag}{\rm diag}
\newcommand{\odiag}{\mbox{\scriptsize off-diag}}
\newcommand{\ptau}{\partial_\tau}
\newcommand{\pth}{\partial_\theta}
\newcommand{\ptaup}{\partial_{\tau'}}
\newcommand{\pthp}{\partial_{\theta'}}
\newcommand{\vev}[1]{\langle #1 \rangle}
\newcommand{\Tr}{\mathop{\mathrm{Tr}}}
\makeatletter
\renewcommand\section{\@startsection {section}{1}{\z@}%
		{-3.5ex \@plus -1ex \@minus -.2ex}%
		{2.3ex \@plus.2ex}%
		{\normalfont\large\bfseries}}
\makeatother

\begin{document}

\title{On the quantum matrix string\footnote{
Presented at the 3rd International Sakharov Conference on Physics,
 Moscow, June 2002.} }

\author{{\sc Shozo Uehara}\footnote{e-mail:
 uehara@eken.phys.nagoya-u.ac.jp}~ and {\sc Satoshi
 Yamada}\footnote{e-mail: yamada@eken.phys.nagoya-u.ac.jp}
\vspace{4mm}\\
{\sl Department of Physics, Nagoya University},\\
{\sl Nagoya 464-8602, Japan}}
\date{}

\maketitle
\vspace{-75mm}
\begin{flushright}
DPNU-02-34\\
hep-th/0210261\\
October 2002
\end{flushright}
\vspace{50mm}
\begin{abstract}
We study the behavior of matrix string theory in the strong coupling
region, where it is expected to reduce to discrete light-cone type
IIA superstring.
In the large $N$ limit, the reduction corresponds to the
double-dimensional reduction from wrapped supermembranes on $R^{10}
\times S^1$ to type IIA superstrings on $R^{10}$ in the light-cone
gauge, which is shown classically, however it is not obvious quantum
mechanically.
We analyze the problem in matrix string theory by using the strong
coupling ($1/g$) expansion.
We find that the quantum corrections do not cancel out
at $\mathcal{O}(1/g^2)$.
Detailed calculations can be seen in Ref.\cite{UY}.
\end{abstract}

\section{Introduction}
Supermembrane in eleven dimensions \cite{HLP,BST} plays an important
role to understand the fundamental degrees of freedom in M-theory.
At the classical level, it was shown that the supermembrane
is related to type IIA superstring in ten dimensions by the
double-dimensional reduction \cite{DHIS}.
The procedure is the following: (i) Consider the target space of
$R^{10}\times S^1$. (ii) Set the compactified coordinate (with radius
$L$) proportional to one of the spatial coordinates of the world volume,
which we call $\rho$ coordinate. (iii) Simply ignore the infinite
tower of the Kaluza-Klein (non-zero) modes.
However, it is not obvious whether such a reduction is justified
also in quantum theory.

Sekino and Yoneya analyzed the double-dimensional reduction quantum
mechanically with the light-cone supermembrane action \cite{SY}.
They kept the Kaluza-Klein modes associated with the $\rho$ coordinate
in the wrapped supermembrane theory on the target space $R^{10}\times
S^1$ and integrated them out by using the perturbative expansion with
respect to the radius $L$.
Since the gauge coupling satisfies $g\sim 1/L$ in the wrapped
supermembrane theory, the expansion can be regarded as the strong
coupling expansion.
They calculated the effective action for the zero modes along the
$\rho$ direction to the one-loop order of $O(L^2)$  and found
that the quantum corrections cancel out and the effective action
agrees with the classical (free) action of type IIA superstring except
at the points where the usual string interactions could occur.
As is emphasized in their paper \cite{SY}, however, the strong
coupling expansion does not give a rigorous proof of the quantum
double-dimensional reduction because the propagators are proportional
to the two-dimensional $\delta$-function, $\delta^{(2)}(\xi) \equiv
\delta(\tau)\delta(\sigma)$, which will cause the ultraviolet
divergences of $\delta^{(2)}(0)$ type in loops.
However it is very difficult to find a suitable regularization which
respects symmetries, and hence the strong coupling expansion is not
yet defined rigorously.
In this sense, they gave a formal argument for the vanishing of the
one-loop corrections of $O(L^2)$ by demonstrating that the
coefficients of $\delta^{(2)}(0)$ coming from both bosonic and
fermionic degrees of freedom cancel out.

The purpose of our work is essentially to extend their analysis to the
two-loop order of $O(L^2)$.
However, the naive extension is not straightforward because at the
two-loop level, even the coefficients of the $\delta^{(2)}(0)$ diverge
due to the contribution of the infinite Kaluza-Klein towers.
Thus, we need another regularization for the summation over the
infinite tower of the Kaluza-Klein modes.
We know the matrix regularization of the supermembrane on
$R^{11}$ in the light-cone gauge \cite{dHN} and also that of the
wrapped supermembrane on $R^{10} \times S^1$ in the light-cone gauge
\cite{SY}.
The former is called Matrix theory \cite{BFSS} and the latter is
called matrix string theory \cite{Mot,DVV} which will be a
non-perturbative formulation of light-cone quantized type IIA
superstring theory in the large $N$ limit. Furthermore,
even at finite $N$, Matrix and matrix string theories
are conjectured to be non-perturbative formulations of discrete
light-cone quantized (DLCQ) M-theory and type IIA superstring theory,
respectively \cite{Sus,Sei,Sen}.
Thus, we consider matrix string theory and study whether
the reduction from matrix strings to discrete light-cone type IIA
superstrings is justified quantum mechanically.\footnote{We use a
convention of the light-cone coordinates such that $x^{\pm}=(x^{0}\pm
x^{10})/\sqrt 2$. Furthermore, $x^-$ is compactified on $S^1$ with
radius $R$ in DLCQ.}

\section{From wrapped supermembrane to matrix string}
The supermembrane action on the target space $R^{11}$ \cite{dHN} in
the light-cone gauge is given by
\begin{eqnarray}
  S&=&LT\int d\tau \!\int_0^{2\pi}\!d\sigma d\rho \Biggl[
    \frac{1}{2}(D_{\tau}X^i)^2 -\frac{1}{4L^2}\{ X^i,X^j\}^2\nn
  &&\hspace{20ex} +i\psi^T D_{\tau}\psi +\frac{i}{L}
	\psi^T\gamma^i\{ X^i,\psi\}\Biggr]\label{action1},\\
  &&D_{\tau}  =\ptau  -\frac{1}{L}\{A,~~\},\quad
  \{A,B\} = \partial_{\sigma}A\,\partial_{\rho}B -
	\partial_{\rho}A\, \partial_{\sigma}B ,
\end{eqnarray}
where the indices $i,j$ run through $1,2,\cdots,9$,
the spinor $\psi$ has sixteen real components\footnote{We use the real
and symmetric gamma matrices $\gamma^i$, which
satisfy $\{\gamma^i,\gamma^j\}=2\delta^{ij}$.}
and $T$ is the membrane tension.
At this stage, $L$ is an arbitrary length parameter of no physical
meaning. The action is invariant under the gauge transformation,
\begin{equation}
  \delta A=\ptau \Lambda + \frac{1}{L}\{\Lambda,A\},\quad
  \delta X^i= \frac{1}{L}\{\Lambda,X^i\},\quad
  \delta \psi= \frac{1}{L}\{\Lambda,\psi\}.
\end{equation}
This gauge transformation generates the area-preserving diffeomorphism
on the world volume.
When the spatial surface of the supermembrane has a non-trivial
topology, we have to impose further the global constraints.

Now we consider the wrapped supermembrane theory on the target space
$R^{10}\times S^1$ and discuss the correspondence with matrix string
\cite{SY}. We take the $X^9$ direction as $S^1$ and identify the
radius with the above parameter $L$,
\begin{eqnarray}
	X^9=L\rho +Y.\label{L}
\end{eqnarray}
Thus $L$ has the physical meaning of the radius of the $X^9$ direction
which is regarded as the ``eleventh'' direction in M-theory.
Substituting eq.(\ref{L}) into eq.(\ref{action1}), we obtain the
light-cone gauge supermembrane action on
$R^{10}\times S^1$,
\begin{eqnarray}
  S&=&LT\int d\tau \int_0^{2\pi}d\sigma d\rho \left[
	\frac{1}{2}F_{\tau\sigma}^2 +\frac{1}{2}(D_{\tau}X^k)^2
	-\frac{1}{2}(D_{\sigma}X^k)^2 -\frac{1}{4L^2}\{ X^k,X^l\}^2
	\right.\nn
  &&\hspace{20ex} \left.+i\psi^T D_{\tau}\psi -i\psi^T \gamma^9
	D_{\sigma}\psi +\frac{i}{L}\psi^T\gamma^k\{ X^k,\psi\}
	\right], \label{action}\\
  &&F_{\tau\sigma}=\ptau  Y - \partial_{\sigma}A -
	\frac{1}{L}\{A,Y\},\quad
  D_{\sigma}=\partial_{\sigma}-\frac{1}{L}\{Y,~~\},
\end{eqnarray}
where the indices $k,l$ run through $1,2,\cdots,8$.
This is also an action of the gauge theory of the area-preserving
diffeomorphism, where the gauge coupling $g\sim 1/L$.
In Ref.\cite{SY}, the area-preserving diffeomorphism in
eq.(\ref{action}) was regularized by the finite dimensional group
$U(N)$ and it was shown that the matrix-regularized form of the action
(\ref{action}) agrees with that of matrix string theory,
\begin{eqnarray}
  &&S=LT\int d\tau \int_0^{2\pi}d\theta \,tr\left[
	\frac{1}{2}F_{\tau\theta}^2 +\frac{1}{2}(D_{\tau}X^k)^2
	-\frac{1}{2}(D_{\theta}X^k)^2 +\frac{1}{4L^2}[ X^k,X^l]^2
	\right.\nn
  &&\hspace{20ex} \left.+i\psi^T D_{\tau}\psi -i\psi^T \gamma^9
	D_{\theta}\psi -\frac{1}{L}\psi^T\gamma^k[X^k,\psi]
	\right],\label{action2}\\
  &&F_{\tau\theta}=\ptau  Y - \pth A - \frac{i}{L}[A,Y],~
  D_{\tau}=\ptau -\frac{i}{L}[A,~~],~
 D_{\theta}=\pth -\frac{i}{L}[Y,~~]\,,
\end{eqnarray}
where each element of the matrices is a function of ($\tau,\theta$).
The action (\ref{action2}) is invariant under the $U(N)$ gauge
transformations,
\begin{eqnarray}
  &&\delta A=\ptau  \Lambda + \frac{i}{L}[\Lambda,A],\quad
  \delta Y=\pth  \Lambda + \frac{i}{L}[\Lambda,Y],\nn
  &&\delta X^k= \frac{i}{L}[\Lambda,X^k],\quad
  \delta \psi= \frac{i}{L}[\Lambda,\psi].\label{gauge}
\end{eqnarray}
The zero-modes along the $\rho$ direction in the wrapped supermembrane
are mapped to the diagonal elements of matrix string while the
Kaluza-Klein modes are mapped to the off-diagonal elements \cite{SY}.
Note that in the matrix regularization of the wrapped supermembrane on
$R^{10} \times S^1$, there are no obvious counterparts of the global
constraints, because the (matrix-regularized) Gauss law constraint,
which is derived from eq.(\ref{action2}), cannot be manifestly
interpreted as the integrability condition.

In the classical double-dimensional reduction, the Kaluza-Klein modes
of every field along the $\rho$ direction are set zero.
And then in the $L\to0$ limit, the action (\ref{action}) reduces to
the light-cone type IIA superstring action.
As for the matrix-regularized action (\ref{action2}),
the off-diagonal elements of every matrix are set zero in such a
classical double-dimensional reduction.
Then the action reduces to the DLCQ type IIA superstring action
in the light-cone momentum $p^+=N/R$ sector.
It is expected that the reductions are justified also in quantum
theory, however, it is not so simple \cite{Rus,SY}.
In particular, the quantum double-dimensional reduction of the wrapped
supermembrane was analyzed for the small radius $L$ in \cite{SY},
which corresponds to the strong gauge coupling $g\sim 1/L$ in the
wrapped supermembrane theory and also to the weak string coupling
$g_s\sim L/\sqrt{\alpha'}$ in type IIA superstring theory.
By using the perturbative expansion with respect to $L$, the
Kaluza-Klein modes along the $\rho$ direction were integrated out to
the one-loop order of $O(L^2)$ and it was found that the effective
action for the zero modes agrees with the classical (free) action of
the type IIA superstring except at the interaction points of
perturbative strings.
So far the result is consistent with the expectation that the wrapped
supermembrane theory in the region of small radius $L$ agrees with
the perturbative type IIA superstring theory.
Then we analyze the quantum reduction of the matrix string
(\ref{action2}) to the diagonal elements for small radius $L$
to study whether the effective action for the diagonal matrix elements
agrees with the classical (free) action of the DLCQ type IIA
superstring.

\section{Strong coupling expansion in matrix string
theory\label{SCEinMST}}
First every $N\times N$ hermite matrix in eq.(\ref{action2}) is
decomposed into the diagonal and off-diagonal parts,
\begin{equation}
 A\to a+A,\quad Y\to y+Y,\quad X^k\to x^k+X^k,
	\quad \psi\to \psi+\Psi,\label{decomp}
\end{equation}
where $a,y,x^k$ and $\psi$ are the diagonal and $A,Y,X^k$ and $\Psi$
are the off-diagonal parts of the original matrices, respectively,
which are plugged into (\ref{action2}).
The gauge transformations are also decomposed as
\begin{equation}
  \delta a=\ptau \lambda +\frac{i}{L}[\Lambda,A]_{\diag},\quad
  \delta A=\ptau  \Lambda + \frac{i}{L}([\lambda,A]+[\Lambda,a]
	+[\Lambda,A]_{\odiag}),~\cdots
\end{equation}
where $\lambda$ and $\Lambda$ are the diagonal and off-diagonal
parts of the gauge parameter, respectively.
At this stage, we impose a boundary condition in $\theta$-direction.
Here, for simplicity, we choose such boundary conditions as
to have the $N$ string bits having $p^+=1/R$ for the diagonal matrix
elements,
\begin{equation}
  \phi_a(\theta +2\pi)=\phi_a(\theta).\quad
	(\mbox{for  } \phi=a,y,x^k,\psi)
\end{equation}
As for the off-diagonal matrix elements, we naturally impose
\begin{equation}
  \Phi_{ab}(\theta +2\pi)=\Phi_{ab}(\theta).\quad
	(\mbox{for  } \Phi=A, Y, X^k, \Psi)
\end{equation}

Next we fix the gauge. We choose the following gauge condition,
\begin{equation}
  a=y,\quad\pth Y-\frac{i}{L}[y,Y]
	-\frac{i}{L}[x^k,X^k]+\frac{i}{L}[a,A]
	-\ptau A=0,\label{fixing}
\end{equation}
and we proceed in the Landau gauge. Then we get
\begin{eqnarray}
 T&=&\int \calD y \calD x^k \calD\psi \calD c \calD\bar{c}
	\calD A \calD Y \calD X^k \calD \Psi \calD C \calD \bar{C}
	\calD B\nn
   &&\hspace{2ex}\exp\Biggl[iLT \int d\tau \int_0^{2\pi}
    d\theta\, (\calL^{string}+ \calL^B_0 +L\calL^B_1+L^2\calL^B_2\nn
    &&\hspace{23ex}+\calL^F_0 +L^{1/2}\calL^F_{1/2}
    +L\calL^F_1)\Biggr],\label{PI}\\
  \calL^{string}&=&\Tr\bigg[\frac{1}{2}(\ptau x^k)^2
	- \frac{1}{2}(\pth x^k)^2
	+\frac{1}{2}\left\{(\ptau -\pth)y\right\}^2\nn
  &&\hspace{2cm}+i\bar{c}(\ptau -\pth )c
	+ i\psi^T\ptau  \psi-i
	\psi^T\gamma^9\pth  \psi \bigg],\\
 \calL^B_0&=&\Tr\bigg[-\frac{1}{2}\,[x^k,A]^2+\frac{1}{2}\,[x^k,Y]^2
	+\frac{1}{2}\,[x^k,X^l]^2\nn
  &&\hspace{1ex}-\frac{1}{2}\left([y,Y]+[x^k,X^k]-[y,A]\right)^2\nn
  &&\hspace{1ex}-iB \left([y,Y]+[x^k,X^k]-[y,A]\right)
  +i[x^k,\bar{C}][x^k,C]\bigg],\label{freeb}\\
 \calL^F_0&=&\Tr\left[\Psi^T[y,\Psi]-\Psi^T\gamma^9[y,\Psi]-
		\Psi^T\gamma^k[x^k,\Psi]\,\right],\label{freef}\\
 \calL^F_{1/2}&=&\Tr\left[-2\Psi^T[\psi,A]+2\Psi^T\gamma^9[\psi,Y]
		+2\Psi^T\gamma^k[\psi,X^k]\,\right],\label{1f}\\
 \calL^B_1&=&\Tr\bigg[-i\ptau Y [y,Y]+2i\ptau Y [y,A]
	-i\ptau A [y,Y]\nn
  &&\hspace{4ex}+ 2i\pth A [y,Y]- i\pth A[y,A]-i\pth Y[y,A]\nn
  &&\hspace{4ex}-i\ptau X^k[y,X^k]+2i\ptau X^k[x^k,A]
	-i\ptau A[x^k,X^k]\nn
  &&\hspace{4ex}+i\pth X^k[y,X^k]-2i\pth X^k[x^k,Y]
	+i\pth Y[x^k,X^k]\nn
  &&\hspace{4ex}-[y,Y][A,Y]+[y,A][A,Y]-[y,X^k][A,X^k]\nn
  &&\hspace{4ex}+[x^k,A][A,X^k]+[y,X^k][Y,X^k]-[x^k,Y][Y,X^k]\nn
  &&\hspace{4ex}+[x^k,X^l][X^k,X^l]+B\pth Y-B\ptau A
	-\pth\bar{C}[y,C]\nn
  &&\hspace{4ex}-[y,\bar{C}]\pth C +i[y,\bar{C}][Y,C]
	+i[x^k,\bar{C}][X^k,C]\nn
  &&\hspace{4ex}+ [y,\bar{C}]\,\ptau C -i[y,\bar{C}][A,C]
	+\ptau \bar{C}[y,C]\bigg],\label{1b}\\
 \calL^F_{1}&=&\Tr\Big[\,i\Psi^T\ptau \Psi
		-i\Psi^T\gamma^9\pth \Psi\nn
  &&\hspace{4ex}+\Psi^T[A,\Psi]-\Psi^T\gamma^9[Y,\Psi]
		-\Psi^T\gamma^k[X^k,\Psi]\,\Big],\label{2f}\\
  \calL^B_2&=&\Tr\bigg[\,\frac{1}{2}(\ptau  Y - \pth A)^2
	+\frac{1}{2}(\ptau  X^k)^2 -\frac{1}{2}(\pth  X^k)^2\nn
  &&\hspace{3ex}-i\ptau Y[A,Y] +i\pth A[A,Y]
	-i\ptau X^k[A,X^k]\nn
  &&\hspace{3ex} +i\pth X^k[Y,X^k] -\frac{1}{2}\,[A,Y]^2
	 -\frac{1}{2}\,[A,X^k]^2\nn
  &&\hspace{3ex}+ \frac{1}{2}\,[Y,X^k]^2+ \frac{1}{4}\,[X^k,X^l]^2
	-i\pth \bar{C}\pth C -\pth \bar{C}[Y,C]\nn
  &&\hspace{3ex} +i\ptau \bar{C}\ptau C	+\ptau\bar{C}[A,C]
	+i[\bar{C},A]_{\diag}[C,A]_{\diag}\nn
  &&\hspace{3ex} -i[\bar{C},Y]_{\diag}[C,Y]_{\diag}
  -i[\bar{C},X^k]_{\diag}[C,X^k]_{\diag}\bigg],\label{2b}
\end{eqnarray}
where ($c$,$\bar{c}$) are (ghost, anti-ghost) for the first
condition of (\ref{fixing}), while ($C$,$\bar{C}$,$B$) are (ghost,
anti-ghost, $B$-field) for the second one, respectively,
$a$ has been integrated out by using the Landau
gauge condition for eq.(\ref{fixing}) and some of
the off-diagonal parts have been rescaled as \cite{SY}
\begin{equation}
  A\to LA, \quad Y\to LY,\quad X^k\to LX^k,\quad\Psi\to L^{1/2}\Psi,
  \quad C\to L^2C.
\end{equation}

By using the above action, we perform the perturbative expansion with
respect to $L$ and integrate only the off-diagonal matrix elements;
\begin{eqnarray}
  &&T=\int \calD y \calD x^k \calD\psi \calD c \calD\bar{c}\,\,
	\exp\left(iS_{eff}[y,\,x^k,\,\bar{c},\,c,\,\psi]\right),\\
  &&S_{eff}[y,\,x^k,\,\bar{c},\,c,\,\psi]=\int d\tau \int_0^{2\pi}
	d\theta\, \left(\calL^{string}
	-i\ln Z[y,\,x^k,\,\psi]\right),\label{eff}\\
  &&Z[y,\,x^k,\,\psi]=\int \calD A \calD Y \calD X^k \calD\Psi\calD C
	\calD\bar{C}\calD B\,\exp \,(i\tilde{S})\,,\label{Z}\\
  &&\tilde{S}= \int d\tau \int_0^{2\pi} d\theta\,
	\Big(\calL^B_0 +L\calL^B_1 + L^2\calL^B_2
      +\calL^F_0 +L^{\frac{1}{2}}\calL^F_{1/2} +L\calL^F_1\Big),
\end{eqnarray}
where we have set $LT=1$ and $\xi=(\tau,\theta)$ for brevity.
We regard (\ref{freeb}) and (\ref{freef}) as the
free parts and (\ref{1f})--(\ref{2b}) the interactions.
Then we read off the propagators from the free parts,
\begin{eqnarray}
  \vev{\hat{X}_{ab}^K(\xi)\hat{X}_{ba}^L(\xi')}
    &=&-i\left(\delta^{KL}-\frac{(\hat{x}_a^K
    -\hat{x}_b^K)(\hat{x}_a^L-\hat{x}_b^L)}{(\hat{x}_a-\hat{x}_b)^2}
    \right)\frac{G(\xi,\xi')}{(\hat{x}_a-\hat{x}_b)^2}\,,\label{pro}\\
  \vev{B_{ab}(\xi)Y_{ba}(\xi')} &=&\vev{B_{ab}(\xi)A_{ba}(\xi')} =
	\frac{y_a-y_b}{(x_a-x_b)^2}\,G(\xi,\xi'),\\
  \vev{B_{ab}(\xi)X_{ba}^k(\xi')}&=&
	\frac{x_a^k-x_b^k}{(x_a-x_b)^2}\,G(\xi,\xi'),\\
  \vev{\bar{C}_{ab}(\xi)C_{ba}(\xi')}&=&
	\frac{G(\xi,\xi')}{(x_a-x_b)^2},\\
  \vev{\Psi_{ab}^{\alpha}(\xi)\Psi_{ba}^{\beta}(\xi')}&=& -\frac{i}{2}
	\frac{(y_a-y_b)(I+\gamma^9)_{\alpha\beta}
	+(x_a^k-x_b^k)\gamma^k_{\alpha\beta}}{(x_a-x_b)^2}G(\xi,\xi')\,
	\label{psipsi},
\end{eqnarray}
where $G(\xi,\xi')\equiv \delta^{(2)}(\xi-\xi')$,
$(x_a-x_b)^2\equiv(x_{a}^k-x_b^k)(x_{a}^k-x_b^k)$,
the spinor indices $\alpha,\beta$ run through $1,2,\cdots,16$
and we have introduced the hatted variables $\hat{X}^K$ and
$\hat{x}^K$ $(K=k,9,10\,\ (k = 1,2,\cdots,8))$,
\begin{equation}
  \hat{X}^k=X^k,\quad\hat{X}^9=Y,\quad\hat{X}^{10}=iA,\quad
  \hat{x}^k=x^k,\quad\hat{x}^9=y,\quad\hat{x}^{10}=iy,\label{hatX}
\end{equation}
and $(\hat{x}_a-\hat{x}_b)^2\equiv(\hat{x}_a^K-\hat{x}_b^K)
(\hat{x}_a^K-\hat{x}_b^K)=(x_a-x_b)^2$.
Notice that $(x_{a}-x_b)^2$ (for $a\ne b$,\ $1\leq a,b\leq N$) must be
non-zero in order that the perturbative expansion makes sense since
the propagators are singular at $(x_a-x_b)^2=0$.
We recall that in matrix string theory, the usual string interactions
are described by the exchange of coincident diagonal matrix elements
and hence the perturbative expansion does not make sense even for
small radius $L$ at the interaction points.
Thus, henceforth we ignore the interaction points and integrate out
the off-diagonal matrix elements to get the effective action for the
diagonal matrix elements, which is expected to agree with the
classical (free) action of DLCQ type IIA superstring.

The perturbative calculation is, however, formal as in Ref.\cite{SY}:
The propagators (\ref{pro})-(\ref{psipsi}) are proportional to the
$\delta$-function $G(\xi,\xi')=\delta^{(2)}(\xi-\xi')$ and
the loops suffer from the ultraviolet divergences like
$\delta^{(2)}(0)$.
However, it is very difficult to find a suitable regularization
which respects symmetries.
If we adopt a certain regularization, e.g. cutoff regularization for
large momenta, the regularized $\delta$-function $G_{(r)}(\xi,\xi')$
would not satisfy $f(\xi)G_{(r)}(\xi,\xi')=f(\xi')G_{(r)}(\xi,\xi')$,
which causes an ambiguity of how we choose the arguments of
$(x_a^k-x_b^k)$ and $(y_a-y_b)$, which appear in the propagators
(\ref{pro})-(\ref{psipsi}).
To avoid the ambiguity, henceforth we consider only the configurations
of the diagonal matrix elements in which the differences of arbitrary
two elements $(x_a^k-x_b^k)$, $(y_a-y_b)$ and
$(\psi_a-\psi_b)$ are independent of $\xi$,
although $x_a^k$, $x_b^k$, $y_a$, $y_b$, $\psi_a$ and $\psi_b$
themselves depend on $\xi$, in general.
We have not yet found such a suitable regularization, however,
we give a formal argument about the quantum corrections order by order
in the strong-coupling expansion.
 
(i) $O(L^{0})$:
The lowest order contribution in eq.(\ref{Z}) is the one-loop
determinant of the free action. Actually, the determinant is unity due
to the coincidence between bosonic and fermionic degrees of freedoms.

(ii) $O(L^{1/2})$:
The next contribution in eq.(\ref{Z}) comes from
$\tilde{S}_{1/2}^F=\int \calL_{1/2}^F$.
$\vev{i\tilde{S}_{1/2}^F}$ vanishes because
there is no way to self-contract in $i\tilde{S}_{1/2}^F$.

(iii) $O(L^{1})$:
The $O(L^{1})$ contributions in eq.(\ref{Z}) come from
$\tilde{S}_{1}^B$, $\tilde{S}_{1/2}^F$ and
$\tilde{S}_{1}^F$.
There are tree kinds of contributions,
$\vev{i\tilde{S}_{1}^B}$, $(1/2!) \vev{i\tilde{S}_{1/2}^F
\,i\tilde{S}_{1/2}^F}$ and $\vev{i\tilde{S}_{1}^F}$.
The first one is given by
\begin{eqnarray}
\vev{i\tilde{S}_{1}^B} &=&iL\int d^2 \xi d^2 \xi'
	 \sum_{a,b=1}^N\bigg\{i(y_a-y_b)\ptau
	\vev{Y_{ab}(\xi)Y_{ba}(\xi')}\nn
  &&\hspace{7ex}-i(y_a-y_b)\ptau\vev{Y_{ab}(\xi)A_{ba}(\xi')}
	-i(y_a-y_b)\pth\vev{A_{ab}(\xi)Y_{ba}(\xi')}\nn
  &&\hspace{7ex}+i(y_a-y_b)\pth\vev{A_{ab}(\xi)A_{ba}(\xi')}
	+i(y_a-y_b)\ptau\vev{X_{ab}^k(\xi)X_{ba}^k(\xi')}\nn
  &&\hspace{7ex}-i(x_a^k-x_b^k)\ptau\vev{X_{ab}^k(\xi)A_{ba}(\xi')}
	-i(y_a-y_b)\pth\vev{X_{ab}^k(\xi)X_{ba}^k(\xi')}\nn
  &&\hspace{7ex}+i(x_a^k-x_b^k)\pth\vev{X_{ab}^k(\xi)Y_{ba}(\xi')}
	  -\pth \vev{B_{ab}(\xi)Y_{ba}(\xi')}\nn
  &&\hspace{7ex}+\ptau\vev{B_{ab}(\xi)A_{ba}(\xi')}
	+2(y_a-y_b)\,\pth\vev{\bar{C}_{ab}(\xi)C_{ba}(\xi')}\nn
  &&\hspace{18ex}-2(y_a-y_b)\,\ptau
	\vev{\bar{C}_{ab}(\xi)C_{ba}(\xi')}\bigg\}
	\,\delta^{(2)}(\xi-\xi').
\end{eqnarray}
{}From eqs.(\ref{pro})-(\ref{psipsi}), we see that the quantity in the
braces is antisymmetric in $a$ and $b$ and hence
$\vev{i\tilde{S}_{1}^B}$ is zero. Similarly we find that
both $\vev{i\tilde{S}_{1/2}^F\,i\tilde{S}_{1/2}^F}$ and
$\vev{i\tilde{S}_{1}^F}$ are zero.
Note that in order to show that the quantum correction of $O(L)$ is
zero, we have never used the fact that $G(\xi,\xi')$
is a $\delta$-function.
Hence it would hold even if $G(\xi,\xi')$ is some
regularized $\delta$-function.

(iv) $O(L^{3/2})$:
There are tree kinds of contributions,
\[
 \frac{1}{2!}\vev{i\tilde{S}_{1}^B\,i\tilde{S}_{1/2}^F},\quad
  \frac{1}{3!}\vev{i\tilde{S}_{1/2}^F\,i\tilde{S}_{1/2}^F\,
	i\tilde{S}_{1/2}^F},\quad
  \frac{1}{2!}\vev{i\tilde{S}_{1}^F\,i\tilde{S}_{1/2}^F}.
\]
Each of them is zero because there is no way of
contraction, respectively.

(v) $O(L^{2})$:
The contributions are
\begin{eqnarray*}
&&\vev{i\tilde{S}_{2}^B},\quad
\frac{1}{2!}\vev{i\tilde{S}_{1}^B i\tilde{S}_{1}^B},\quad
\frac{1}{2!}\vev{i\tilde{S}_{1}^F \,i\tilde{S}_{1}^F},\quad
\frac{1}{2!}\vev{i\tilde{S}_1^B i\tilde{S}_1^F},\\
&&\frac{1}{3!}\vev{i\tilde{S}_{1/2}^F i\tilde{S}_{1/2}^F
	i\tilde{S}_{1}^B},\quad
  \frac{1}{3!}\vev{i\tilde{S}_{1/2}^F i\tilde{S}_{1/2}^F
	i\tilde{S}_{1}^F},\quad
  \frac{1}{4!}\vev{ i\tilde{S}_{1/2}^F  i\tilde{S}_{1/2}^F
    i\tilde{S}_{1/2}^F  i\tilde{S}_{1/2}^F}.
\end{eqnarray*}
The last three terms contain fermionic diagonal
elements $\psi_a$ and they each vanish due to the anti-commutativity
of the Grassmann variables $\psi_a$. Also it is easy to see that
there is no contribution from $(1/2!)\vev{i\tilde{S}_1^B
i\tilde{S}_1^F}$.
Then $(1/2!)\vev{i\tilde{S}_{1}^B  i\tilde{S}_{1}^B}$,
$\vev{i\tilde{S}_{2}^B}$ and
$(1/2!)\vev{i\tilde{S}_{1}^F \,i\tilde{S}_{1}^F}$ are to be considered
below.

(v-1) One-loop:
The one-loop contributions coming from
$(1/2!)\vev{i\tilde{S}_{1}^B  i\tilde{S}_{1}^B}$,
$\vev{i\tilde{S}_{2}^B}$ and
$(1/2!)\vev{i\tilde{S}_{1}^F \,i\tilde{S}_{1}^F}$
are referred to as
$(1/2!)\vev{i\tilde{S}_{1}^B  i\tilde{S}_{1}^B}^{(1)}$,
$\vev{i\tilde{S}_{2}^B}^{(1)}$ and
$(1/2!)\vev{i\tilde{S}_{1}^F \,i\tilde{S}_{1}^F}^{(1)}$,
respectively. They are given by
\begin{eqnarray}
  &&\frac{1}{2!}\,\vev{i\tilde{S}_{1}^B
	i\tilde{S}_{1}^B}^{(1)}=L^2\int d^2 \xi d^2 \xi'
	\sum_{a\ne b}\Biggl[-\frac{17(y_a-y_b)^2}{\{(x_a-x_b)^2\}^2}
	\ptau\pthp G(\xi,\xi')\,G(\xi,\xi')\nn
  &&\hspace{6ex} +\left(
	\frac{1}{(x_a-x_b)^2}+\frac{17}{2}\,\frac{(y_a-y_b)^2}
	{\{(x_a-x_b)^2\}^2}\right)
	\ptau\ptaup G(\xi,\xi')\,G(\xi,\xi')\nn
  &&\hspace{6ex}-\left(
	\frac{1}{(x_a-x_b)^2}-\frac{17}{2}\,
	\frac{(y_a-y_b)^2}{\{(x_a-x_b)^2\}^2}\right)
	\pth\pthp G(\xi,\xi')\,G(\xi,\xi')\Biggr],\label{1-loop1}\\
  &&\vev{i\tilde{S}_{2}^B}^{(1)}= L^2\int d^2 \xi d^2
	\xi'\sum_{a\ne b}\Bigg[\frac{(y_a-y_b)^2}{\{(x_a-x_b)^2\}^2}\,
	\ptau \pthp G(\xi,\xi')\,\delta^{(2)}(\xi-\xi')\nn
  &&\hspace{6ex}\left(\frac{3}{(x_a-x_b)^2}
	-\frac{1}{2}\frac{(y_a-y_b)^2}{\{(x_a-x_b)^2\}^2}\right)
	\ptau \ptaup G(\xi,\xi')\,\delta^{(2)}(\xi-\xi')\nn
  &&\hspace{6ex}
	-\left(\frac{3}{(x_a-x_b)^2}
	+\frac{1}{2}\frac{(y_a-y_b)^2}{\{(x_a-x_b)^2\}^2}\right)
	\pth\pthp G(\xi,\xi')\,\delta^{(2)}(\xi-\xi')\Bigg],
	\label{1-loop2}\\
  &&\frac{1}{2!}\vev{i\tilde{S}_{1}^F
	i\tilde{S}_{1}^F}^{(1)} =L^2\int d^2 \xi d^2 \xi'
	\sum_{a \ne b}\Bigg[\frac{16(y_a-y_b)^2}{\{(x_a-x_b)^2\}^2}\,
	\ptau \pthp G(\xi,\xi')\,G(\xi,\xi')\nn
  &&\hspace{6ex}+\left(\frac{-4}{(x_a-x_b)^2}
	-\frac{8(y_a-y_b)^2}{\{(x_a-x_b)^2\}^2}\right)
	\ptau \ptaup G(\xi,\xi')\,G(\xi,\xi')\nn
  &&\hspace{6ex}+\left(\frac{4}{(x_a-x_b)^2}
	-\frac{8(y_a-y_b)^2}{\{(x_a-x_b)^2\}^2}\right)
	\pth \pthp G(\xi,\xi')\,G(\xi,\xi')\Bigg]\label{1-loop3}.
\end{eqnarray}
Note that we have never used the fact that $G(\xi,\xi')$ is the
$\delta$-function in deriving eqs.(\ref{1-loop1})-(\ref{1-loop3}).
Thus they are expected to be unaltered even if we adopt a certain
regularization.
At this stage we first use the fact that $G(\xi,\xi')$ is the
$\delta$-function and it is shown that they are canceled,
\footnote{This result in matrix string theory is essentially
the same as the one in the wrapped supermembrane theory \cite{SY}.
In Ref.\cite{SY}, however, the zero-mode gauge field $a$ is restricted
to be zero by hand, while we have just fixed the gauge ($a=y$) and
added the corresponding FP-ghost part following the standard procedure
\cite{KU}. In this sense the configuration of $a$ is not restricted in
our calculations.}
\begin{equation}
 \frac{1}{2!}\vev{i\tilde{S}_{1}^B  i\tilde{S}_{1}^B}^{(1)}
  +\vev{i\tilde{S}_{2}^B}^{(1)}
  + \frac{1}{2!}\vev{i\tilde{S}_{1}^F\,i\tilde{S}_{1}^F}^{(1)}=0.
\end{equation}

(v-2) Two-loop:
The two-loop contributions, which are not calculated in Ref.\cite{SY},
are referred to as $(1/2!)\vev{i\tilde{S}_{1}^B
i\tilde{S}_{1}^B}^{(2)}$, $\vev{i\tilde{S}_{2}^B}^{(2)}$ and
$(1/2!) \vev{i\tilde{S}_{1}^F \,i\tilde{S}_{1}^F}^{(2)}$,
respectively. We obtain
\begin{eqnarray}
  &&\frac{1}{2!}\vev{i\tilde{S}_{1}^B
    i\tilde{S}_{1}^B}^{(2)} =-i L^2\int d^2 \xi d^2 \xi'
    \sum_{a\ne b,\,b\ne c,\,c\ne a}\Biggl\{\frac{33}{2}
        \frac{1}{(x_a-x_b)^2(x_b-x_c)^2}\nn
   &&\hspace{22ex}-16\,\frac{\{(x_a^k-x_b^k)(x_c^k-x_a^k)\}^2}
	{(x_a-x_b)^2(x_b-x_c)^2\{(x_c-x_a)^2\}^2}\nn
   &&\hspace{22ex}-\frac{1}{2}\,
    \frac{\{(x_a^k-x_b^k)(x_b^k-x_c^k)\}^2}{\{(x_a-x_b)^2(x_b-x_c)^2\}^2}
	\Biggr\}\,(G(\xi,\xi'))^3,\label{1'}\\
  &&\vev{i\tilde{S}_{2}^B}^{(2)}=iL^2\int d^2 \xi d^2\xi'
	\Biggl[\sum_{a\ne b,\, b\ne c,\,c\ne a}
	\left\{\frac{73}{2}\frac{1}{(x_a-x_b)^2(x_b-x_c)^2}\right.\nn
  &&\hspace{18ex}-\frac{1}{2}\frac{\{(x_a^k-x_b^k)(x_b^k-x_c^k)\}^2}
	{\{(x_a-x_b)^2(x_b-x_c)^2\}^2}\Biggr\}
	+54 \sum_{a\ne b} \frac{1}{\{(x_a-x_b)^2\}^2}\Biggr]\nn
  &&\hspace{20ex}
	\times (G(\xi,\xi'))^2\,\delta^{(2)}(\xi-\xi'),\label{2'}\\
  &&\frac{1}{2!} \vev{i\tilde{S}_{1}^F
	\,i\tilde{S}_{1}^F}^{(2)} =-iL^2\int d^2 \xi d^2 \xi'
	\sum_{a\ne b,\,b\ne c,\,c\ne a}
	\bigg\{20\,\frac{1}{(x_a-x_b)^2(x_b-x_c)^2}\nn
   &&\hspace{15ex}+16\,\frac{\{(x_a^k-x_b^k)(x_c^k-x_a^k)\}^2}
	{(x_a-x_b)^2(x_b-x_c)^2\{(x_c-x_a)^2\}^2}
	\bigg\}(G(\xi,\xi'))^3. \label{3'}
\end{eqnarray}
Note that in calculating eqs.(\ref{1'})--(\ref{3'}) we have never used
the fact that $G(\xi,\xi')$ is the $\delta$-function and hence
eqs.(\ref{1'})--(\ref{3'}) are expected to be unaltered
even if we adopt a certain regularization.
At this stage we first use the fact that $G(\xi,\xi')$ is the
$\delta$-function and sum up eqs.(\ref{1'})--(\ref{3'}) to get
\begin{eqnarray}
 &&\frac{1}{2!}\vev{i\tilde{S}_{1}^B
    i\tilde{S}_{1}^B}^{(2)}+\vev{i\tilde{S}_{2}^B}^{(2)}
	  +\frac{1}{2!} \vev{i\tilde{S}_{1}^F
	\,i\tilde{S}_{1}^F}^{(2)}\nn
 &&\hspace{1ex}=iL^2\int d^2 \xi d^2\xi'\sum_{a\ne b}
	\frac{54}{\{(x_a-x_b)^2\}^2}\,(G(\xi,\xi'))^3.\label{eq:2Lsum}
\end{eqnarray}
\textit{The two-loop quantum corrections at $O(L^2)$ do not cancel
out!}  One comment is in order:  The remaining term is exactly
that of the second summation in eq.(\ref{2'}).
If we assume that the differences of the diagonal elements can be
estimated as $(x_a^k-x_b^k)\sim O(N^{\alpha})$\footnote{According to
the correspondence of a long string in matrix string theory
with the wrapped supermembrane, $\alpha=-1$ for $|a-b|\ll N$
\cite{SY}.} with some common constant $\alpha$ for large $N$,
we will see that the terms canceled in eq.(\ref{eq:2Lsum})
behave as $\sum_{a\ne b,\,b\ne c,\,c\ne a} (x_a^k-x_b^k)^{-2}
(x_b^k-x_c^k)^{-2} \sim O(N^{3-4\alpha})$, while the remaining term
behaves as $\sum_{a\ne b} (x_a^k-x_b^k)^{-4}\sim O(N^{2-4\alpha})$.
In this sense, we could say that only the leading terms in the large
$N$ are canceled in the two-loop quantum corrections to the
classical string action at $O(L^2)$.

\section{Conclusion}
We have studied in matrix string theory whether the reduction to
the diagonal elements of the matrices is justified quantum
mechanically. We have seen that the quantum corrections do not cancel
out at $O(L^2)$. We should note, however, that no suitable
regularization for the divergences of $\delta^{(2)}(0)$ type is found
so far, and hence we have only studied a mechanism of cancellation of
quantum corrections.
In fact, we have found that at the two-loop level of $O(L^2)$,
the sub-leading term in the large $N$ comes only from the bosonic
degrees of freedom and cannot be canceled out.
Even if we find a suitable regularization, such a structure seems to
be unaltered and hence our result will be unchanged.

Finally, we comment on the global constraints in the wrapped
supermembrane theory.
Such constraints should be taken into account in the calculations of
the quantum double-dimensional reduction.\footnote{The global
constraints are not considered in the calculations \cite{SY}.}
In matrix string theory, however, there are no counterparts of such
constraints.
In particular, in the standard derivation of matrix string theory,
they do not appear naturally.
However we can show that the global constraint does not alter our
result \cite{UY2}.

\vspace{\baselineskip}
\noindent
\textbf{Acknowledgments.}
One of the authors (SU) would like to thank the organizers of the
``Third International Sakharov Conference on Physics'' for the kind
invitation. The work of SU is supported in part by the Grant-in-Aid
for Scientific Research No.13135212.


\end{document}